\documentclass[prl,twocolumn]{revtex4} 
\usepackage{epsfig}

\begin{document}

\title{Thermopower of a Kondo spin-correlated quantum dot}

\author{R. Scheibner, H.\ Buhmann, D. Reuter$^\dag$, M.N. Kiselev$^*$ and L.W. Molenkamp}

\affiliation{Physikalisches Institut (EP3) and $^*$Institut f{\"u}r Theoretische Physik, Universit\"at W\"urzburg,
97074 W\"urzburg, Germany\\$^\dag$Lehrstuhl f{\"u}r Festk{\"o}rperphysik, Ruhr-Universit{\"a}t Bochum, 44801 Bochum, Germany}

\date{\today}

\begin{abstract}
The thermopower of a Kondo-correlated gate-defined quantum dot is studied using a current heating technique. In
the presence of spin correlations the thermopower shows a clear deviation from the semiclassical Mott relation
between thermopower and conductivity. The strong thermopower signal indicates a significant asymmetry in the
spectral density of states of the Kondo resonance with respect to the Fermi energies of the reservoirs. The
observed behavior can be explained within the framework of an Anderson--impurity model.
\\



\end{abstract}

\maketitle


The Kondo effect due to magnetic impurity scattering in metals is a well known and widely studied phenomenon
\cite{Hewson:97}. The effect has recently received much renewed attention since it was demonstrated
\cite{Goldhaber-Gordon:98,Cronenwett:98} that the Kondo effect can significantly influence transport through a
semiconductor quantum dot (QD). In a gate defined QD, the electronic states can be controlled externally, which
allows experimenters to address many questions concerning Kondo physics \cite{vanderWiel:2000} that were
previously inaccessible. As yet unexplored are the thermoelectric properties of a QD in the presence of Kondo
correlations. This is unfortunate, since these properties often yield valuable additional information concerning
transport phenomena.  For example, the thermopower (TP) $S$ is related to the average energy $\langle E \rangle$
with respect to the Fermi energy $E_{\rm F}$ of the particles contributing to the transport  by
\cite{Ziman:60,Chaikin:76}
\begin{equation}\label{TPeqn}
S \equiv - \lim_{\Delta T \rightarrow 0} \left.\frac{V_{\rm T}}{\Delta T}\right|_{I=0} =-\frac{\langle E
\rangle}{eT} \text{ ,}
\end{equation}
where $V_{\rm T}$ is the thermovoltage, $\Delta T$ the applied temperature difference, and $e$ the elementary
charge. $S$ is therefore a direct measure of the weighted spectral density of states in a correlated system with
respect to the Fermi energy $E_F$. Moreover, the TP can be used to determine the spin-entropy flux accompanying
electron transport, which sets boundaries on the operating regime of spin-based quantum computers \cite{Loss:98}.
Spin entropy was recently shown to be the origin of the giant thermoelectric power of layered cobalt oxides
\cite{Wang:03}, and one may anticipate similarly large effects for single quantum dots.

In this Letter we present TP measurements on a lateral QD in the presence of Kondo correlations in comparison with
results obtained for the weak coupling Coulomb blockade (CB) regime. The experiments show a clear breakdown of
transport electron-hole symmetry in the vicinity of Kondo spin-correlations, accompanied by deviations from the
semiclassical Mott relation \cite{Ziman:60},
\begin{equation}\label{Motteqn}
 S_{\rm Mott}= - \frac{\pi^2}{3}\frac{k^2T}{e} \frac{\partial \ln
G(E)}{\partial E}\text{,}
\end{equation}
where $k_B$ is Boltzmann's constant, and $G(E)$ is the energy-dependent conductance of the QD.
%
%
\begin{figure}
\centering
\includegraphics [width = 7.6 cm] {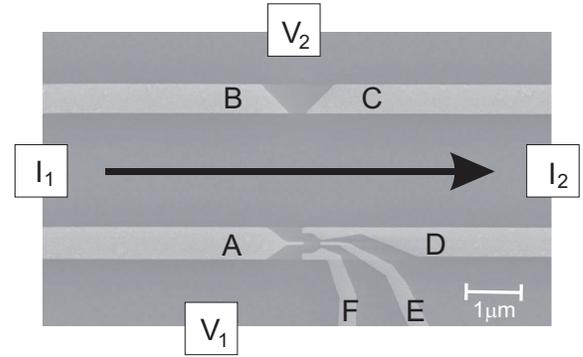}
\caption{SEM photograph of the quantum dot and the central part of the 20~$\mu$m long heating channel. The labels
$\text{V}_{1,2}\text{ and } \text{I}_{1,2}$ indicate the 2DEG areas with ohmic contacts. }
\end{figure}

Figure~1 shows an SEM-photograph of the sample structure. This structure is fabricated by electron-beam
lithography on a (Al,Ga)As/GaAs heterostructure containing a two dimensional electron gas (2DEG) with carrier
density $n_s = 2.3\times 10^{15}$~m$^{-2}$ and mobility $\mu=100$~m$^2$/(Vs). Gates A, D, E, and F form the QD and
gates A, B, C, and D are the boundaries of the electron-heating channel, which is 20~$\mu$m long and 2~$\mu$m
wide. The QD has a lithographical diameter of approximately 250~nm and has a design \cite{Ciorga:00} that allows
great flexibility in the number of electron on the dot. For the present experiment, the barriers are adjusted such
that the number of electrons can be varied conveniently (i.e., by changing only the voltage $V_{\rm E}$ applied to
plunger gate E) between 20 and 40, as can be deduced from the magnetic-field evolution of the CB peaks. The sample
is mounted in a top loading dilution cryostat with a base temperature $T_{\rm bath}$ below 50~mK.
%
%
\begin{figure}
\centering
\includegraphics [width = 8.2 cm] {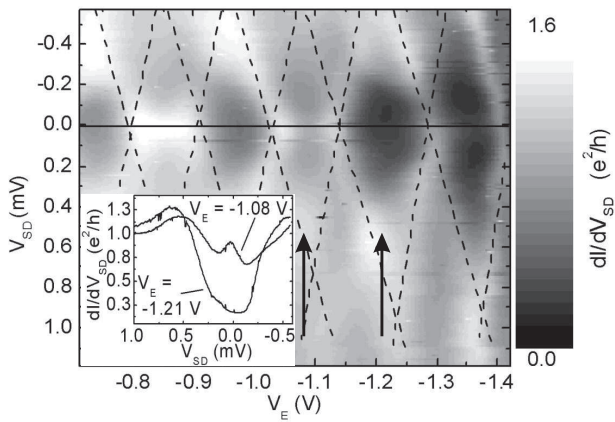}
\caption{Grey-scale plot of the differential conductance as a function of the QD potential $(\propto V_{\rm E})$
and the externally applied bias voltage across the QD ($V_{\rm SD}$). Alternating regimes of low and high
conductance are observed between each successive conductance peaks within the CB diamonds (dashed lines). Inset:
Bias-voltage ($V_{\rm SD}$) depending traces of the differential conductivity of the dot for $V_{\rm E} = -1.08$~V
and $V_{\rm E}=-1.21$~V [indicated by arrows].}
\end{figure}
For conductance measurements gates B and C are grounded. Figure~2 shows a grey-scale plot of the differential
conductance of the QD as a function of its potential ($\propto V_{\rm E}$) and the bias voltage across the dot
($V_{\rm SD}$). Along the zero bias line [$V_{\rm SD}=0$, solid line in Fig.\ 2], alternating regimes of low and
high conductance in between two successive conductance peaks are observed in the gate-voltage range $-0.7 < V_{\rm
E} < -1.2$~V. A $V_{\rm SD}$-dependent trace [inset of Fig.\ 2] for $V_{\rm E} = -1.08$~V (high conductance
between two main CB peaks) shows a clear zero bias resonance, in contrast to a $V_{\rm SD}$ trace taken in the
adjacent conductance valley ($V_{\rm E} = -1.21$~V).

Such a zero bias resonance is characteristic for a Kondo correlated system \cite{Hewson:97}. In the following, we
assign the regions around $V_{\rm E} \approx -0.85$~V and -1.08~V as voltage ranges where spin correlations modify
the transport characteristics. For $V_{\rm E}$ more negative than -1.2~V the coupling between the QD electrons and
the surrounding 2DEG decreases and the Kondo correlations are suppressed. From temperature dependent measurements
and the analysis of the CB diamonds \cite{Kouwenhoven:97} we deduce charging energies varying from $E_{\rm C} =
$~0.7~meV to $E_{\rm C} = $~1.5~meV and corresponding intrinsic level widths of $\Gamma \sim$~0.35~meV to $\Gamma
\sim$~0.15~meV in the regions of strong and weak coupling of the dot to the reservoirs, respectively.

Experimentally, the main problem in performing TP measurements on nanostructures is how to obtain a sizable
temperature difference across a very small device. Here, we use a current heating technique \cite{Molenkamp:90,
Molenkamp:92}, that previously has been applied to measure the TP of metallic QDs \cite{Moeller:98}. A small
ac-heating current is passed through the channel defined by gates A, B, C, and D. Energy dissipation in the
channel, results in a local heating of the electron gas. Due to the small electron-phonon coupling in (Al,Ga)As
2DEGs at low temperatures, hot electrons can only dissipate their excess energy to the lattice in the wide 2DEG
area behind the channel exit, while rapid electron-electron scattering ensures thermalization of the electrons in
the channel to a temperature $T_{\rm ch}$ which is higher than the lattice temperature $T_{\rm L}$. Hence, the QD
is embedded between the hot electron reservoir in the channel (with electron temperature $T_{\rm el}=T_{\rm ch}$)
and the cold surrounding 2DEG ($T_{\rm el} = T_{\rm L}$). The constant temperature difference of $\Delta T =
T_{\rm ch}- T_{\rm L}$ across the QD can be adjusted via the current through the channel. For the experiments
presented in Figs.\ 3 and 4(a), we use a heating current of $I_{\rm H}= 4.2$~nA. This results in $\Delta T \approx
10$~mK, where $\Delta T$ is calibrated by making use of the quantized TP of the quantum point contacts formed by
gates B and C (QPC$_{\rm BC}$) and A and B \cite{Molenkamp:92}. (For this calibration the QD, gates E and F are
set to ground.)

The thermovoltage across the QD is measured as the voltage difference $V_{\rm T} = V_1 - V_2$ (see Fig.\ 1).
$V_{\rm T}$ contains the TP of the QD ($S_{\rm QD}$) as well as that of QPC$_{\rm BC}$ ($S_{\rm QPC}$). For
convenience, the TP of QPC$_{\rm BC}$ is adjusted to zero by setting its conductance at the center of a
conductance plateau. Figure 3 presents our results on $V_{\rm T}$ as a function of gate voltage $V_{\rm E}$ (lower
panel) together with the corresponding conductance curve (upper panel). As discussed above, for gate voltages less
negative than $-1.1$~V a strong coupling of the QD to the leads results in Kondo correlations at $V_{\rm E}
\approx -0.72$~V and $-0.95$~V, indicated by the horizontal bars underneath the top (conductance) panel. [Note
that these values have shifted with respect to those in Fig.~2 because of the additional negative gate voltages on
gates B and C] In the bottom (thermovoltage) panel, a striking difference between the behavior of $V_{\rm T}$ in
the spin correlated regime as compared with the weak coupling regime is directly conspicuous.

In the weak coupling regime the line shape of the thermovoltage has negative and positive contributions in the
vicinity of a conductance peak. As shown in Ref.~\cite{Turek:01}, this behavior results from contributions of two
different transport mechanisms: (1) the linear increase in thermovoltage at the center of a conductance peak is
characteristic of sequential tunneling, while (2) its rapid fall-off between two conductance peaks is a
consequence of cotunneling processes. In this regime, the thermovoltage line-shape follows qualitatively the
negative parametric derivative of the conductance data, as described by Mott's relation, Eq.~(\ref{Motteqn})
\cite{Mott,Scheibner:04}. The grey line in the bottom panel of Fig.~3 shows the behavior of the thermopower as
expected from Mott's relation.
%
%
\begin{figure}
\centering
\includegraphics [width = 8.6 cm] {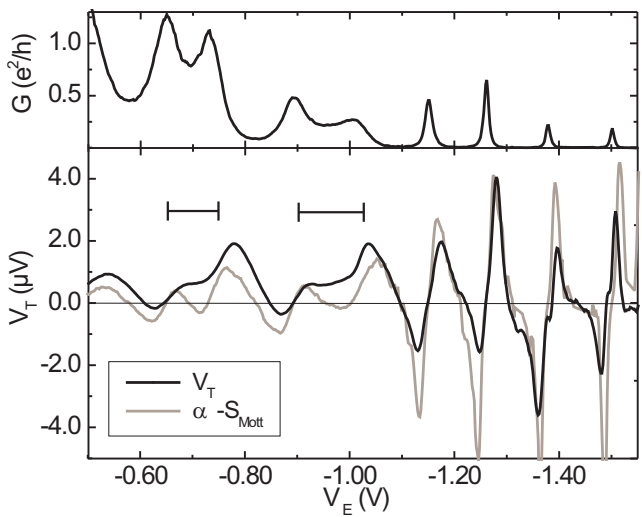}
\caption{Comparison of the thermovoltage measurement ($V_{\rm T}$), the conductance (G), and the calculated
thermopower $S_{\rm Mott}$ as expected from the Mott-relation [Eq.\ (\ref{Motteqn}), where $E\propto (-e V_{\rm
E})$] as a function of applied gate voltage $V_{\rm E}$. The horizontal bars indicate the regions of spin
correlations within the gate voltage range of strong coupling of the quantum dot to the leads ($V_{\rm
E}>-1.1\text{V}$).}
\end{figure}

The striking experimental result for the strong coupling regime is now that in the presence of spin correlations
at $V_{\rm E}= -0.72$~V  and $-0.95$~V, $V_{\rm T}$ exhibits {\it only positive} values, while when these
correlations are absent ($V_{\rm E}= -0.62$~V  and $-0.83$~V) the usual negative $V_{\rm T}$ shows up in between
the Coulomb resonances. A comparison with the semiclassically expected $S_{\rm Mott}$ [c.f. Eq. (2)] in Fig.\ 3
shows at the additional contributions at $V_{\rm E}= -0.72$~V  and $-0.95$~V an amount of ca. $0.5$ $\mu$V, which
thus cannot originate from single particle effects. Clearly, spin correlations are a prime candidate for
explaining the occurrence of these extra contributions.

Further information can be obtained from the dependencies of these effects on lattice temperature $T_{\rm L}$,
which are shown in Fig.\ 4. [Note that these experiments were done on the same sample but for a different cooling
cycle, where the regime of spin correlations was observed for a different adjustment of the voltages applied to
the QD gates.] In Fig.\ 4(a) we plot the temperature dependence of the TP anomaly. We observe that the additional
contributions to the TP are suppressed at higher temperatures. As a consequence, the expected valley reappears
between the main CB resonances. For comparison we show in Figs.\ 4(b) and (c) the temperature evolution of the
bias($V_{\rm SD}$)-dependent differential conductance and the zero bias conductance of the QD. We observe that the
zero bias resonance in these curves disappears on a similar temperature scale as the extra features in the TP,
establishing a very strong indication that these features are related to spin correlations.
%
%
\begin{figure}
\centering
\includegraphics [width = 8.2 cm] {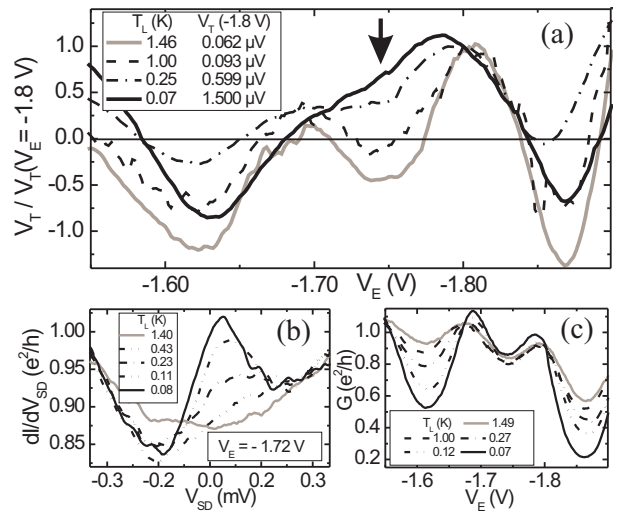}
\caption{(a) Thermovoltage signal for various lattice temperatures at constant $\Delta T$. The curves are
normalized to the value at $V_{\rm E}=$~$-1.8$~V. At high temperatures the spin contribution to the thermopower
between two CB peaks decreases and the oscillating CB substructure of the thermopower reappears (indicated by the
arrow). (b,c) Temperature dependent (differential) conductance of the zero bias resonance as a function of bias
voltage ($V_{\rm SD}$) and QD potential ($V_{\rm E}$).}
\end{figure}
%
%

The anomalous behavior of the TP in the spin-correlation regime points to an asymmetry in the electron-hole
transport, which can be understood as follows: In a TP measurement the electron heating channel has a higher
temperature and therefore more empty (occupied) states below (above) $E_F$ than the colder surrounding 2DEG. When
a QD level has an energy $\varepsilon_{\rm QD}$ slightly higher than the Fermi level $E_F$ of the reservoirs ($E_F
< \varepsilon_{\rm QD} < E_{\rm F} + k_BT$), the dominating transport process is electron-like: an electron first
tunnels onto the QD and then out again into the other lead. The average energy $\langle E \rangle$ of the charges
moving from the hot reservoir to the cold reservoir is positive with respect to $E_F$ \cite{Turek:01}.
Correspondingly, if $E_F
> \varepsilon_{\rm QD} > E_F - k_BT$, $\langle E \rangle$ is
negative referred to $E_F$ and the charge transport can be described as a hole-like process. Hence, one can
directly deduce the relative position of $\varepsilon_{QD}$ to $E_F$ from the sign of the TP signal, as well as
infer whether the contributing transport process is electron- or hole-like.

In order to explain our observation of a positive thermovoltage signal, the above reasoning directly implies that
the spectral density of the correlated state on the QD must have its weighted maximum above $E_F$ in the leads.
From the experimental characterization of our QD, we find that $\varepsilon_{\rm QD} /\Gamma > -3$ in the gate
voltage region where we observe spin correlations [c.f. Fig.\ 3 ($V_{\rm E}>-1.2 \text{ V}$)]. By describing the
spin correlated QD in terms of an asymmetric Anderson model (i.e. $\varepsilon_{QD}\neq-E_{\rm C}/2$
\cite{epsilon}) \cite{Hewson:97,Costi:94,Boese:01,Dong:02}, this implies that we are in the mixed-valence regime.
In this limit, the spectral density of the hybridized state has its maximum above $E_F$ \cite{Costi:94,Hewson:97}.
This is because strong coupling to the reservoirs leads to significant charge valence fluctuations, i.e. a
delocalization of the QD charge, which results in an overall asymmetry between electron and hole like transport.
One thus expects enhanced positive contributions to the thermovoltage \cite{Costi:94}. This is in contrast to
numerical calculations for an ideal Kondo-QD system, where one expects both positive and negative contributions to
the TP \cite{Dong:02}.

We note here that Kondo QDs reported on experimentally very often operate in the mixed valence regime
\cite{Goldhaber-Gordon:98,Cronenwett:98}. Because of the insensitivity of zero bias conductance measurements to
the exact location of the spectral density of the hybridized state, this notion may have escaped general
attention. The TP is in contrast very sensitive to this location \cite{Boese:01}, which is why we observe positive
contributions to the TP. Further deviations of the experiment from the single level Anderson model occur due to
the intrinsic level broadening of the QD states, which even leads to an overlap of the states across the CB gap,
resulting in a non-zero conductance in the CB valleys without spin correlations at $V_{\rm E}>-1.1\text{V}$ (c.f.
Fig.~3).

Finally, we may verify the nature of the entropy flux driving the thermovoltage. In principle, one expects
contributions from both spin and orbital degrees of freedom to the entropy flux \cite{ChaikinKwak:79}. However,
for strongly correlated systems the spin degrees of freedom should dominate the TP \cite{Chaikin:76}. In the spin
correlated regime, the spin degeneracy on the QD opens a channel for spin entropy transport between the
reservoirs. According to Heikes' formula \cite{Chaikin:76}, $S=-\sigma/e $, where $\sigma$ is the entropy change
during the transport of a single particle through the QD, we estimate a spin entropy contribution to the TP of
$S_s = (k_B/e) \ln 2 \approx -60$~$\mu$V/K. From the measurements (Fig.~3) we find an additional thermovoltage at
the resonances between the CB peaks at $V_{\rm E}= -0.72$~V and $-0.95$~V of $V_{\rm T}\sim 0.5$ $\mu$V which
yields a TP of $S\approx -50$~$\mu$V/K using Eq.~(\ref{TPeqn}) and $\Delta T \approx 10$~mK. This value agrees
very well with the expected value for $S_s$ and shows that the effect we measure is strongly spin-entropy driven.
This observation provides independent and direct evidence of the correlated character of the transport through the
Kondo QD. Previously, such direct information has only been obtained in bulk materials, such as 1-dimensional
organic salts \cite{ChaikinKwak:79} and, recently, in cobalt oxides \cite{Wang:03}.

In summary, we have measured the TP of a small QD. The TP exhibits an additional contribution due to spin entropy
when the dot is tuned into the spin correlated (Kondo) regime. The comparison of the TP in the pure CB regime with
the TP in the spin-correlated regime reflects a lifting of the symmetry in the electron-hole transport. The
semiclassical Mott relation between the TP and the conductance fails to describe this asymmetry, which is due to
the many-particle nature of the correlation induced resonance. The measurements agree with theoretical
considerations \cite{Costi:94,Boese:01} addressing the evolution of the TP of a Kondo correlated system as a
function of QD energy. The presented results confirm that the spectral density of states for the Kondo-resonance
of a dot can be explained in the framework of an Anderson impurity model in the presence of charge fluctuations.
Our observation demonstrates that it is possible to create correlated thermoelectric transport in man-made
nanostructures, where the experimenter has a close control of the exact transport conditions. Future detailed
studies will address the quenching of the spin correlation in in-plane magnetic fields, in analogy to
Ref.~\cite{Wang:03}, as well as the magnitude and scaling behavior of the spin correlation contribution to the TP
in comparison with results of renormalization group calculations.

We wish to thank C. Gould, R. Fazio, J. Martinek and M. Sindel for valuable discussions, and T. Borzenko and M.
K{\"o}nig for the device fabrication. This work has been done with the financial support of the DFG (Mo 771/5-2) and
ONR (04PR03936-00).


\begin{thebibliography}{99}
\bibitem[15]{Mott} Note that Mott's relation is not quantitatively valid in the
sequential tunneling regime. See also ref. \protect\cite{Scheibner:04}.
\bibitem[20]{epsilon} Note that using the notation of the Anderson model as
described in Refs. \cite{Hewson:97} and \cite{Costi:94}, $\varepsilon_{\rm QD}$ varies between 0 and $-E_{\rm
C}/2$ during the CB oscillations.
\end{thebibliography}
\end{document}